\documentclass[10pt,letterpaper]{article}
\usepackage{opex3}

\begin{document}
\title{High-Q, low index-contrast polymeric photonic crystal nanobeam cavities}
\author{Qimin Quan$^{1}$$^*$, Ian B. Burgess$^{1}$, Sindy K. Y. Tang$^{1,2}$, Daniel L. Floyd$^{1}$ and Marko Loncar$^{1}$$^{**}$}

\address{$^{1}$ School of Engineering and Applied Science, Harvard
University, Cambridge, MA 02138
\\
$^{2}$Wyss Institute for Biologically Inspired Engineering, Harvard University, Cambridge, MA 02138}
\email{$^*$quan@fas.harvard.edu}
\email{$^{**}$loncar@seas.harvard.edu}

\begin{abstract}
We present the design, fabrication and characterization of high-\emph{Q} (\emph{Q}=36,000) polymeric photonic crystal nanobeam cavities made of two polymers that have an ultra-low index contrast (ratio=1.15) and observed thermo-optical bistability at hundred microwatt power level. Due to the extended evanescent field and small mode volumes, polymeric nanobeam cavities are ideal platform for ultra-sensitive biochemical sensing. We demonstrate that these sensors have figures of merit (FOM=9190) two orders of magnitude greater than surface plasmon resonance based sensors, and outperform the commercial Biacore$^{\mathrm{TM}}$ sensors. The demonstration of high-Q cavity in low-index-contrast polymers can open up versatile applications using a broad range of functional and flexible polymeric materials.
\end{abstract}


\bibliography{amsplain}

\section{Introduction}
The use of polymeric materials in the fabrication of micro- to nano-scale devices has rapidly increased in the past decade. Examples of successful applications can be found in broadband communications\cite{Ma02}, large screen and flexible displays\cite{choi08}, solar cells,\cite{Na08,chen09} and biomedical sensors\cite{Ghezzi11,Broz10} to name a few. However, one area where the realization of all-polymer devices has proved challenging is in the development of high quality factor (defined as $Q=\omega_0 \frac{\mathrm{Energy\, stored}}{\mathrm{Power\,\, loss}}$) nano-scale optical resonators. In particular, it has been a challenge to achieve high-$Q$ photonic crystal (PhC) cavities made of polymeric materials due to the low index contrast between available materials. While there has been some recent demonstrations of ring resonators\cite{chao02,dalton02,yariv03}, microtoroids\cite{armani04} and photonic crystal cavities\cite{Khan11}-\cite{murshidy10} in relatively low-index-contrast platforms ($n_{\mathrm{cav}}/n_{\mathrm{bg}}\sim1.5$), achieving simultaneously high \emph{Q}-factors and small mode volumes (defined as $V=\int \epsilon |\mathbf{E}|^2 \mathrm{d}V/[\epsilon|\mathbf{E}|^2]_{\mathrm{max}}$) in the ultra-low index-contrast regime has remained elusive. Here, we present the design, fabrication and characterization of all-polymer photonic crystal nanobeam cavities, fabricated in the electron-beam resist ZEP520 ($n=1.54$) on a substrate of the spin-on fluoropolymer CYTOP ($n=1.34$). Despite having an extraordinarily small index-contrast ($n_{\mathrm{cav}}/n_{\mathrm{bg}}=1.15$) we achieved \emph{Q}s as high as  \emph{Q}=36,000 - an order of magnitude higher than the state of the art for $n_{\mathrm{cav}}/n_{\mathrm{bg}}=1.46$ \cite{gong10}. Furthermore, nanobeam cavities have much smaller mode volumes compared to the polymeric ring resonators with the same \emph{Q}-factors and made of the same materials. Due to the high \emph{Q}-factors and small mode volumes we observe optical bistability in the polymeric cavities at hundred microwatt power levels.  Furthermore, the greater extension of the evanescent field in our low-index-contrast platform allows polymeric nanobeam cavity-modes to display high sensitivity to the background refractive index. As such we show that polymer nanobeam cavities, as biomedical sensors, have figure of merit (FOM=9190), two orders of magnitude greater than the surface plasmon resonance (SPR) based sensors and outperform commercially avilable Biacore$^{\mathrm{TM}}$ SPR sensors. In addition, the simple fabrication process of polymeric cavities potentially eliminates the need of clean-room facilities and can be easily scaled up for mass production.

\section{High-Q Photonic Crystal Cavity in Ultra-Low Index-Contrast Polymeric Materials}
\begin{figure}
\includegraphics[width=12.0cm]{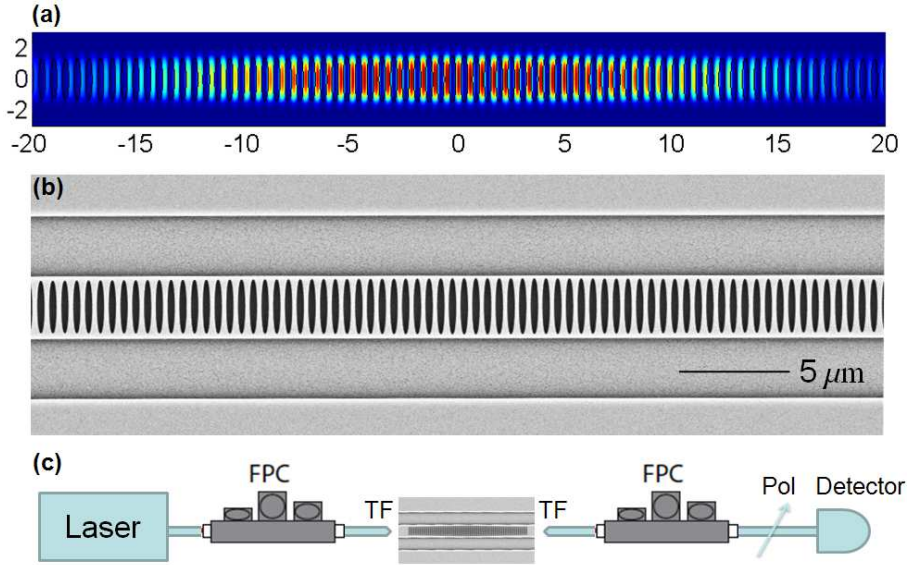}
\caption{(a) Energy density distribution of the cavity mode from FDTD simulation.  (b) Scanning electron micrograph of the polymeric nanobeam cavity. (c) Schematics of the measurement setup. A tunable laser source is coupled to the edge of the chip through a tapered fiber (TF, Ozoptics inc.) after the fiber polarization controller (FPC), and collected from tapered fiber followed by a second FPC and an inline polarizer (Pol) to the detector. The two FPCs are to filter out unwanted TM polarization component that is not in resonance with the cavity.
}\label{sensor}
\end{figure}
\begin{figure}
\setlength{\parindent}{0cm}
\includegraphics[width=12.0cm]{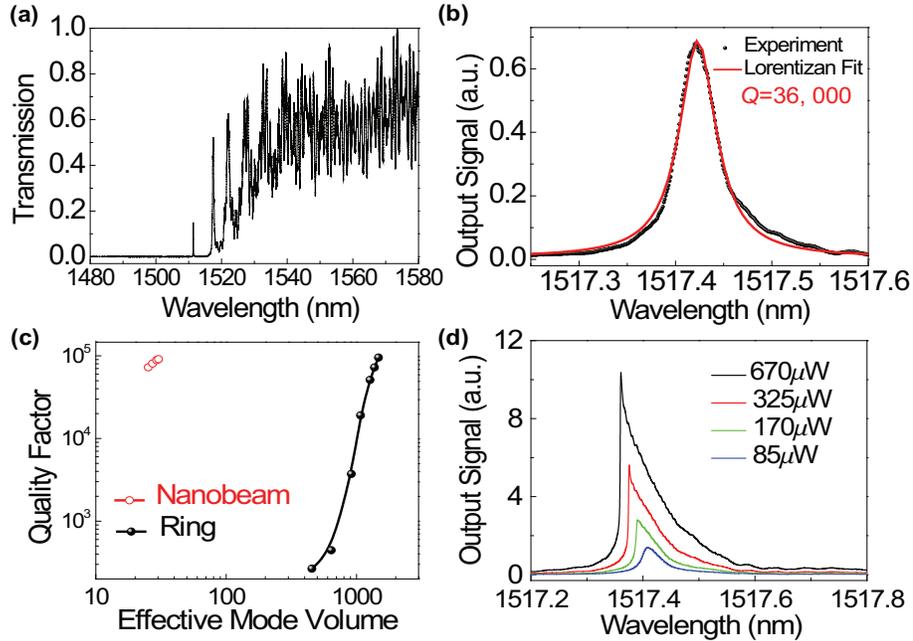}
\caption{(a) Measured optical transmission spectrum of a high-transmission polymeric nanobeam cavity in D$_2$O. The cavity mode has Q=11,000 with a high on-resonance transmission 15\%.
(b) The experimental transmission spectrum of the designed high Q cavity mode. Q of 36,000 is extracted from Lorentzian fit. (c) Comparison of theoretical Q factors and mode volumes between PhC nanobeam cavities and ring resonators with the same waveguide dimension: $3.2\mu m \times 0.5\mu m$, and the same material platform. Mode Volumes are normalized by $(\lambda/n_{\mathrm{ZEP}})^3$. The number of tapered hole pairs in the nanobeam cavities are varied from 45 to 60, with an additional 50 hole pairs on both ends of the tapered section. The radii of the rings are varied from 25$\mu$m to 80$\mu$m. (d) Transmission spectra of the cavity mode at different input powers, showing optical bistability. The power levels indicated
in the legend correspond to the powers coupled into the on-chip waveguide. The laser wavelength was swept from shorter to longer wavelengths across the cavity resonance.
}\label{spectrum}
\end{figure}

The nanobeam cavity geometry consists of a ridge waveguide perforated with gratings of elliptical holes (Figure \ref{sensor}). In our design, we considered a cavity having refractive-index $n_{\mathrm{cav}}=1.54$ (e.g. ZEP520) surrounded by a background index $n_{\mathrm{bg}}=1.34$, which is suitable for biosensing applications. The cavity was designed using the deterministic high-\emph{Q} design method that we previously introduced\cite{quan10,quan11}. The distances between neighboring holes are kept the same as 550nm. There are in total 100 grating sections on both sides. The first 50 grating sections are 50 ellipses whose dimensions linearly decrease from the center to both ends of the 3.2$\mu m$-wide and 500nm-thick waveguide, followed by 50 ellipses that have the same dimensions. In the linearly decreasing section, the major axes of the ellipses decrease from 1.44$\mu$m to 1.22$\mu$m, and the minor axes decrease from 165nm to 140nm. The reason for choosing elliptical shape instead of circular shape is because elliptical grating sections have larger bandgap and higher reflectivity to confine the optical mode. At current level of index contrast between the cavity and the background, a symmetric structure is essential to minimize coupling into substrate modes. A symmetric structure can be fabricated simply by adding a capping layer of CYTOP on top of the on-substrate cavity. However, we note that the refractive index of the CYTOP layer also well matches the refractive index of many common liquids (e.g. water). Thus an equivalent optical structure is formed by immersing the cavity in water instead of adding a polymer capping layer. Therefore, we used the water cladding geometry in our measurements as such a geometry is ideal for optofluidic and sensing applications. We determined the theoretical mode profile and \emph{Q} of the cavity mode using the finite-difference time-domain (FDTD) method (Lumerical Solutions.). The mode profile is shown in Figure \ref{sensor}(a), with a simulation \emph{Q}=86,000. The device was fabricated directly via electron-beam lithography in the positive-tone ZEP520 (Zeon Corp.) resist ($n=1.54$) supported by a $3.2\mu$m layer of the low-index fluoropolymer CYTOP (AGC inc.) ($n=1.34$) on a silicon chip. The CYTOP layer was deposited by multiple spin-coating steps. The polymer layers were sufficiently thin that no charging effects were observed using a 100kV electron-beam (Elionix Inc.), even in the absence of an added conductive layer. Figure \ref{sensor}(b) shows the scanning electron microscope (SEM) image of the center region of a nanobeam cavity.

A schematic of the measurement setup is shown in Figure \ref{sensor}(c). A tunable telecom laser source was coupled to the edge of the chip through a tapered fiber (TF, Ozoptics inc.) after a fiber polarization controller (FPC). Light was then collected from tapered fiber followed by a second FPC and an inline polarizer (Pol) to the detector. The two FPCs filter out the unwanted TM polarization component that is not in resonance with the cavity. We measured the cavities submerged in D$_2$O, which has negligible absorption in the telecom frequency range. Figure \ref{spectrum} (a) shows the full spectrum of one of the cavities. The highest \emph{Q} we obtained was 36,000 in $\mathrm{D}_{2}\mathrm{O}$, as shown in Figure \ref{spectrum}(b). Not only is the index ratio of our system much lower (index ratio=1.15 v.s index ratio$\geq$1.45 in all other cases), the \emph{Q}-factor is an order of magnitude higher than previously demonstrated low index PhC cavities suspended in air \cite{gong10}-\cite{murshidy10}.

\section{Small Mode Volumes}
Small mode volumes are a significant advantage of PhC nanobeam cavities over the whispering gallery mode (WGM) cavities (e.g ring resonators, microspheres etc.) since the strength of light and matter interaction scales inversely with $V$. A general trade off holds that higher Q-factors require larger mode volumes\cite{quan11}. Maintaining high \emph{Q}s in low-index contrast cavities will increase the size of the ring resonators and PhC cavities. However, we found that the mode volumes required to achieve a given $Q$ are much larger for ring resonators than for our nanobeam PhC cavities. We illustrated this by comparing the mode volumes of nanobeam cavities with ring resonators that have the same cross sections with the nanobeam cavity waveguide (i. e. $3.2\mu m \times 0.5\mu m$ cross section), and made of the same materials (ZEP on CYTOP immersed in D$_2$O). Figure \ref{spectrum}(c) shows the relationship between \emph{Q}s and mode volumes for the ring resonators with varying radii and for the nanobeam cavities with different number of tapered hole pairs (i.e. varying lengths). Notably, the ring resonators required to achieve the \emph{measured} \emph{Q}s of our nanobeam cavities have about 50 times larger mode volume, even \emph{theoretically}.

\section{Optical Bistability in Polymeric Cavities}
Due to both high \emph{Q}s and the small $V$s, we were able to observe optical bistability\cite{winful79} in our nanobeam cavity. Optical bistability due to the thermo-optic effect (i.e. change of refractive index due to heat) has been widely observed in silicon cavities\cite{quan10,almeida04,noda06,notomi09}, and most recently observed in polymer ring resonators by Ling et. al. \cite{ling11}. In contrast to silicon case, polymer has a negative thermo-optic coefficient ($\mathrm{d}n/\mathrm{d}\mathrm{T}$) \cite{Ma02} causing the resonance of the cavity to blue-shift with increasing power (Figure \ref{spectrum}(d)). Quantitatively, the heat (in unit of W) that is converted from optical energy ($U$) due to the absorption of polymer can be obtained from:
$h=\omega U/Q_{\mathrm{abs}}=2\sqrt{T}P_{\mathrm{in}}Q_\mathrm{total}/Q_{\mathrm{abs}}$, where $\omega$ is the cavity resonance frequency, $T$ is the on-resonance transmission and $P_{\mathrm{in}}$ is the input optical power. Thermal equilibrium is reached through
conduction and convection processes. The thermal resistance of the cavity relates the heat
generated in the cavity to the temperature rise at the center of the
cavity by $\delta \mathrm{T}=h\cdot R_{\mathrm{th}}$. Therefore the resonance shift due to the thermo-optic effect can be expressed as:
\begin{equation}
\delta \lambda=2 \sqrt{T}Q_\mathrm{total}/Q_{\mathrm{abs}} \frac{\mathrm{d}\lambda} {\mathrm{d}n}\frac{\mathrm{d}n}{\mathrm{dT}} R_{\mathrm{th}} P_{\mathrm{in}}
\end{equation}
The thermo-optic coefficient for ZEP 520 is expected to be $\frac{\mathrm{d}n}{\mathrm{dT}}\sim-10^{-4}$\cite{Ma02}. Using the thermal conductivity $\sigma_{\mathrm{th}}=0.16$W$\mathrm{m}^{-1}\mathrm{K}^{-1}$\cite{ZEPtcon} and heat capacity $C_{p}\sim1.5Jg^{-1}\mathrm{K}^{-1}$, the thermal resistance of the cavity structure was estimated to be $R_{\mathrm{th}}=4.8\times10^4 \mathrm{K} \mathrm{W}^{-1}$ from finite-element simulation (Comsol)\cite{notomi09}. Neglecting all the other heating effects, as well as higher order nonlinear effects that could change the index of the polymer, our experimentally observed resonance shift of 20pm at 170$\mu$W input power indicates $Q_{\mathrm{abs}}\sim10^6$ for linear absorption in the ZEP 520 layer, in agreement with the literature~\cite{PMMAabs,ZEPabs}.

\section{Ultra-Sensitive Glucose Sensor}
\begin{figure}
\setlength{\parindent}{0cm}
\includegraphics[width=12cm]{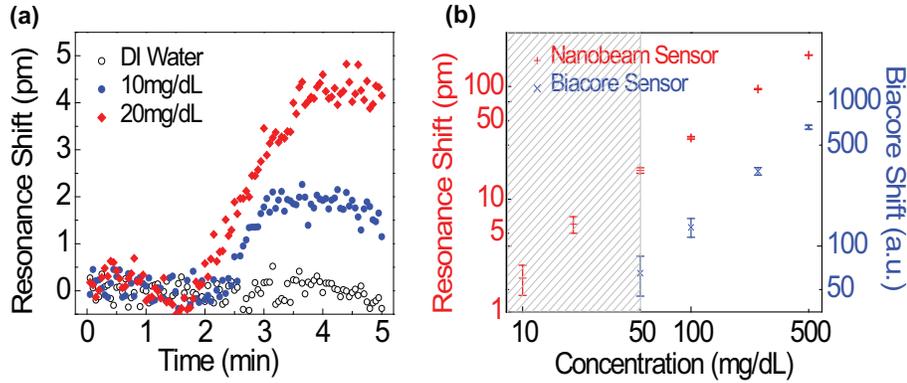}
\caption{ (a) The real-time response of the resonance shift in response to the infusion of pure DI water, and glucose solutions with concentrations of 10mg/dL and 20mg/dL. (b) Equilibrated resonance shifts in different
concentrations of glucose solutions in DI water measured by the nanobeam sensor and Biacore$^{\mathrm{TM}}$ 3000 instrument respectively. The Biacore chip was functionalized with (11-Mercaptoundecyl)tetra(ethylene glycol) to prevent adsorption of molecules on the gold surface to insure the measured responses are due to the bulk glucose index change. }\label{glucose}
\end{figure}
Due to the low index contrast, the evanescent field of the polymeric cavity is more extended into the surrounding medium than the widely used silicon based nanophotonic sensors. As a result, the resonance shift of a polymeric cavity in response to the bulk refractive index change in the surrounding media is larger. Furthermore, our cavity design is optimized for optofluidic integration, with index-matching between the substrate and the fluid layer. To illustrate the potential for such an all-polymer platform for label-free sensing\cite{fan08}, we measured the shift in the cavity resonance in response to varying glucose concentration in water. The glucose solution was delivered to the cavity via a fluidic channel that was integrated on top of the chip. The channel was made of polydimethylsiloxane (PDMS, Sylgard 184, Dow Corning) with two millimeter-diameter holes on both ends as
inlet and outlet. The channel walls were painted with a CYTOP layer to maintain index-matching in the waveguide. Fluid injection was kept at a constant rate of 50$\mu$L/min. The time response to the lowest reproducibly detectable concentration (10mg/dL) is shown in Figure \ref{glucose}(a), along with the null response to pure DI water and the response to 20mg/dL glucose solution. The resonance shifts of glucose solutions at different concentrations are shown in Figure \ref{glucose}(b). Between each measurement, pure DI water was injected to the cavity and the cavity resonance shifted back to its original value. The response of the sensor to the concentration of the glucose solution exhibits excellent linearity covering the whole range of clinically relevant levels. Figure \ref{glucose}(b) further shows a direct comparison of our nanobeam sensor with the well-known commercial label-free biosensor: Biacore$^{\mathrm{TM}}$ instrument. Notably, nanobeam sensors demonstrated about 5 times improvement in the detection limit. The wavelength shift per refractive index unit (RIU) for the nanobeam sensor is $S=386$nm. This is about 4 times larger than similar PhC structures made of silicon, and is about about 2 times larger than ring resonators\cite{white08}. The figure of merit ($\mathrm{FOM}=SQ/\lambda=9190$) is orders of magnitude larger than surface plasmon resonance based sensors (typically 8-23 \cite{anker08}) and metamaterial sensors (330 \cite{kabashin09}). Given the $<$1pm fluctuation in resolving the cavity resonance (Figure \ref{glucose}(a) DI water curve), our sensor can detect a minimum refractive index change of $2\times 10^{-6}$RIU.

\section{Conclusion}
Due to high radiation losses in the low index contrast systems, the demonstration of high-\emph{Q}, small-\emph{V} photonic crystal cavities has been elusive. Here we experimentally demonstrated a photonic crystal cavity with \emph{Q}=36,000 in polymeric materials possessing ultra-low index-contrast. These cavities have much smaller mode volumes than the previously demonstrated polymeric WGM cavities. In addition, due to the extended optical mode profile, we demonstrated high sensitivity and linearity of the polymeric nanobeam sensor in response to different glucose concentrations in water and outperformed the commercial Biacore$^{\mathrm{TM}}$ instrument. The ability to realize high-\emph{Q}/\emph{V} nanophotonic resonators in a non-suspended all-polymer platform makes these devices accessible to a broad range of materials as well as simple and scalable fabrication techniques such as interference lithography, imprint lithography, and replica molding. We expect our results to stimulate a new set of applications by combining functional nanostructures with new materials, such as nonlinear optics, flexible LEDs, optofluidics.

\section{Acknowledgments}
The authors acknowledge fruitful discussions with Parag B. Deotare, Yinan Zhang and Jennifer T. Choy. IBB acknowledges support from NSERC (Canada) through the PGS-D program. This work is supported in part by NSEC at Harvard and AFOSR award FA9550-09-1-0669. Fabrication of the nanobeam cavities was performed at Harvard's Center for Nanoscale Systems.

\end{document}